\documentclass[aps, prd, superscriptaddress, nofootinbib, preprintnumbers, showpacs]{revtex4}
\usepackage{graphicx}
\usepackage{amsmath}
\usepackage[usenames]{color}
\preprint{UWO-TH-07/12}

\usepackage{allpurpose}
\begin{document}


\title{The Derivative Expansion of the Effective Action and the
 Renormalization Group Equation}

\author{F.A. Chishtie}
\email{fachisht@uwo.ca} \affiliation{ Department of Applied
Mathematics, University of Western Ontario, London, Ontario N6A 5B7,
Canada }
\author{Junji Jia}
\email{jjia5@uwo.ca} \affiliation{ Department of Applied
Mathematics, University of Western Ontario, London, Ontario N6A 5B7,
Canada }
\author{D.G.C. McKeon}
\email{dgmckeo2@uwo.ca} \affiliation{ Department of Applied
Mathematics, University of Western Ontario, London, Ontario N6A 5B7,
Canada }

\pacs{11.10.Hi}

\date{\today}

\begin{abstract} The perturbative evaluation of the effective action
can be expanded in powers of derivatives of the external field. We
apply the renormalization group equation to the term in the
effective action that is second order in the derivatives of the
external field and all orders in a constant external field,
considering both massless scalar $\phi_4^4$ model and massless
scalar electrodynamics. A so-called ``on shell'' renormalization
scheme permits one to express this ``kinetic term'' for the scalar
field entirely in terms of the renormalization group functions
appropriate for this scheme. These renormalization group functions
can be related to those associated with minimal subtraction.
\end{abstract}

\maketitle

\section{Introduction}

In the massless $\phi_4^4$ model whose classical action is \be
S_{\mbox{cl}}=\int\dfx\lsb\frac{1}{2}(\dsubmu\phi)(\dsupmu\phi)
-\frac{\lambda}{4!}\phi^4\rsb \label{phi4act},\ee the effective
Lagrangian in the presence of a background field $\phi_{\mbox{cl}}$
can be expressed in powers of the derivatives of $\phi_{\mbox{cl}}$
so that \cite{1} \be
\calcl_{\mbox{eff}}=-V(\phi_{\mbox{cl}}(0))+\frac{1}{2}Z(\phi_{\mbox{cl}}(0))
\lb\dsubmu\phi_{\mbox{cl}}(0)\dsupmu\phi_{\mbox{cl}}(0)\rb+\cdots
.\label{effa}\ee (Henceforth, $\phi_{\mbox{cl}}$ will be simply
denoted by $\phi$.) The effective potential $V$ can be computed
using either diagrammatic \cite{1} or functional methods
\cite{2,3,4}; in both cases divergences arise which require a
renormalization. Consequently, $V$ has explicitly dependence on a
renormalization scale parameter $\mu$. Since \mmu is unphysical,
there must be implicit dependence of $V$ on \mmu through \mlambda
and $\phi$ so that $\displaystyle \mu\frac{dV}{d\mu}=0$. This leads
to the ``renormalization group'' (RG) equation. Solving this
equation perturbatively allows one to find in closed form the sum of
all the so-called ``leading-log'' (LL), ``next-to-leading-log''
(NLL) etc. contributions to $V$. Applying this technique in the
massless standard model leads to a prediction of a Higgs mass of
approximately 220 Gev \cite{5,6}. Furthermore, the on shell
renormalization condition \be \frac{d^4V(\phi=\mu)}{d\phi^4}=\lambda
\label{vcd} \ee has been shown to fix $V$ entirely in terms of the
RG functions associated with this scheme \cite{7}. This is also true
in massless scalar electrodynamics \cite{8}. The renormalization
group has also been applied to analyzing $V$ in Refs.
\cite{18,19,20,21,21p5}.

The RG equation can also be used to extract information about the
kinetic term $Z(\phi)$ occurring in equation \refer{effa}. In this
paper, we follow the approach of Refs. \cite{7,8} to show how
$Z(\phi)$ is in fact completely fixed by the RG functions in much
the same manner that \mcv is determined. No Feynman diagrams need to
be evaluated in the course of determining $Z$. Again, we consider
the $\phi_4^4$ model of eq. \refer{phi4act} and massless scalar
electrodynamics. The need to perform explicit perturbative
calculation of $Z$ through evaluation of Feynman diagrams (as in
Refs. \cite{12,13,14}) is completely circumvented.

The function $Z(\phi)$ is of particular interest, as if \mv is the
value of \mphi that minimizes $V$, so that \be V'(v)=0, \ee then the
radiatively induced mass of the scalar is given by \cite{1} \be
M^2=V''(v)/Z(v). \label{massdef} \ee In the discussion of the
radiatively induced Higgs mass in the standard model in Ref.
\cite{5,6}, $Z(\phi)$ was taken to be close to its classical value
of one.

\section{$Z(\phi)$ in the $\phi_4^4$ model}\label{secz}
Applying the condition \be \mu\frac{d}{d\mu}\lb
Z(\phi)\dsubmu\phi\dsupmu\phi\rb =0 \ee to the kinetic term in the
effective Lagrangian of eq. \refer{effa} results in \be \lsb
\mu\frac{\partial}{\partial
\mu}+\beta(\lambda)\frac{\partial}{\partial
\lambda}+\gamma(\lambda)\phi\frac{\partial}{\partial
\phi}+2\gamma(\lambda)\rsb Z(\lambda,\phi,\mu)=0\label{zrg},\ee
where in eq. \refer{zrg} \bea
\beta(\lambda)\aeq\mu\frac{d\lambda}{d\mu}\\
\gamma(\lambda)\aeq\frac{\mu}{\phi}\frac{d\phi}{d\mu}\eea and the
dependence of $Z$ on \mlambda and \mmu as well as \mphi has been
explicitly noted.

Divergences which arise in the computation of $Z$ require imposition
of a renormalization condition. The condition we chose is \be
Z(\lambda,\phi=\mu,\mu)=1 .\label{zcd}\ee This is an ``on shell''
condition, much the same as the condition on \mcv given in eq.
\refer{vcd} \cite{1}.

The general form of $\beta(\lambda)$, $\gamma(\lambda)$ and
$Z(\lambda,\phi,\mu)$ when the renormalization condition of eq.
\refer{zcd} is applied is \bea \beta(\lambda)\aeq
\sum_{n=2}^{\infty}b_n\lambda^n \label{bexp} \\
\gamma(\lambda)\aeq \sum_{n=1}^{\infty} g_n\lambda^n\label{gexp}\eea
and \be Z(\lambda,\phi,\mu)=\sum_{n=0}^{\infty}\sum_{m=0}^{n}
T_{nm}\lambda^nL^m,\label{zexp}\ee where $L=\ln\lb\phi^2/\mu^2\rb$.
There are several ways of rearranging the sum in eq. \refer{zexp},
one being the ``vertical'' reorganization so that \be
Z=\sum_{m=0}^{\infty}A_m(\lambda)L^m\label{zvre}\ee where \be
A_m(\lambda)=\sum_{n=0}^{\infty}T_{n+m,m}\lambda^{n+m}\ee another
being the ``diagonal'' reorganization \be
Z=\sum_{m=0}^{\infty}\lambda^mS_m(\lambda L)\label{zdre}\ee where
\be S_m(\lambda L)=\sum_{n=0}^{\infty}T_{n+m,m}(\lambda
L)^n.\label{sdef}\ee The function $S_m(\lambda L)$ is the N$^m$LL
contribution to $Z$.

We now will show that the RG equations when applied to $Z$ written
in the form of eq. \refer{zvre} allows one to express $A_m(\lambda)$
in terms of $A_0(\lambda)$. Substitution of eq. \refer{zvre} into
eq. \refer{zrg} results in \be \sum_{m=0}^{\infty}\lsb\lb
-2+2\gamma(\lambda)\rb mL^{m-1}+\lb
\beta(\lambda)\frac{d}{d\lambda}+2\gamma(\lambda)\rb L^m\rsb
A_m(\lambda)=0, \ee which shows that at each order of $L$ \be
A_m(\lambda)=\frac{1}{2m}\lb
\hatbeta(\lambda)\frac{d}{d\lambda}+2\hatgamma(\lambda)\rb
A_{m-1}(\lambda)\label{subz}\ee where $\hatbeta=\beta/(1-\gamma)$
and $\hatgamma =\gamma/(1-\gamma)$.

If now we define \be A_m(\lambda)=\exp\lb
-2\int_{\lambda_0}^{\lambda}\frac{\hatgamma(x)}{\hatbeta(x)}dx\rb
B_m(\lambda),\ee and \be \eta(\lambda)=2\int_{\lambda_0}^{\lambda}
\frac{dx}{\hatbeta(x)}, \ee then eq. \refer{subz} becomes \bea
B_{m}(\lambda)\aeq
\frac{1}{m}\frac{d}{d\eta} B_{m-1}(\lambda(\eta)) \nonumber \\
\aeq \frac{1}{m!}\frac{d^m}{d\eta^m}B_0(\lambda(\eta)).\eea The sum
in eq. \refer{zvre} now becomes
 \bea Z\aeq
\sum_{m=0}^{\infty} \exp\lb
-2\int_{\lambda_0}^{\lambda}\frac{\hatgamma(x)}{\hatbeta(x)}dx\rb\frac{L^m}{m!}\frac{d^m}{d\eta^m}B_0(\lambda(\eta))\\
\mbox{or\hspace{1cm}}&&\nonumber\\
\aeq\exp\lb -2\int_{\lambda(\eta+L)}^{\lambda}
\frac{\hatgamma(x)}{\hatbeta(x)}dx\rb
A_0(\lambda(\eta(\lambda)+L)),\label{a0arb}\eea showing that $Z$
itself can be expressed in terms of $A_0$ and the RG equations
\mbeta and $\gamma$.

If now we impose the renormalization conditions of eq. \refer{zcd},
then as $L=0$ when $\phi^2=\mu^2$ and $\lambda(\eta)=\lambda$, we
see that eq. \refer{a0arb} becomes \be A_0(\lambda)=1 .\ee With
$A_0(\lambda)$ having been so fixed, \mcz is seen now to be
determined solely by the RG functions \mbeta and \mgamma evaluated
in the renormalization scheme of eqs. \refer{vcd} and \refer{zcd}.

It is also convenient to examine the consequence of substituting eq.
\refer{zdre} into \refer{zrg}, along with expansions of eqs.
\refer{bexp} and \refer{gexp}. This results in \bea &&\lsb -2+2\lb
g_1\lambda +g_2\lambda^2+\cdots\rb\rsb  \lsb \lambda S_0'+\lambda^2
S_1'+\lambda^3 S_2'+\cdots\rsb \nonumber \\
&+& \lsb b_2\lambda^2+b_3\lambda^3+\cdots\rsb\lsb \lb S_1+2\lambda
S_2+3\lambda^2 S_3+\cdots \rb +\frac{\xi}{\lambda}\lb S_0'+\lambda
S_1'+\lambda^2 S_2'+\cdots\rb\rsb\nonumber\\
&+&2\lsb g_1\lambda+g_2\lambda^2+\cdots\rsb\lsb S_0+\lambda S_1
+\lambda^2 S_2+\cdots \rsb=0. \label{zdsub}\eea Here, $S_m$ is
evaluated at $\xi\equiv\lambda L$. At order $\lambda^{m+1}$, eq.
\refer{zdsub} generates a differential equation expressing
$S_m(\xi)$ in terms of $S_{m-1}(\xi), \cdots, S_0(\xi)$ as the
following \be \sum_{i=0}^m\lsb\lb 2g_{m-i}+b_{m+2-i}\xi\rb
S_i'(\xi)+\lb 2g_{m+1-i}+ib_{m+2-i}\xi\rb
S_i(\xi)\rsb=0,\label{eqm}\ee where we have defined a constant
$g_0\equiv -1$.

We now impose the renormalization condition of eq. \refer{zcd}; this
combined with eq. \refer{zdre} shows that \be 1=\sum_{m=0}^{\infty}
\lambda^m S_m(0) \ee so that $S_m(0)=\delta_{m0}. $ This serves as
the boundary condition to the differential equations \refer{eqm} for
$S_m(\xi)$.

As $g_1=0$ \cite{11}, eqs. \refer{eqm} have the following solution
\be S_m(\xi)= \frac{1}{b_2^m}\lsb a_{m0}+\sum_{i=1}^m\lb
\frac{1}{w^i}\sum_{j=0}^{i-1}a_{mij}(\ln w)^j\rb\rsb
\label{smsol}\ee where $w=1-b_2\xi/2$ and $a_{m0}$ and $a_{mij}$ are
constants in terms of the coefficients $b_n~(n\leq m+1)$ and
$g_n~(n\leq m+1)$ of RG functions $\beta$ and $\gamma$. In
particular, the solutions to $S_0$ to $S_4$ are \bea && S_0(\xi)= 1
\label{s0sol} \\ && S_1(\xi)=\frac{1}{b_2}\lb -2g_2+\frac{2g_2}{w}\rb \label{s1sol}\\
 &&S_2(\xi)= \frac{1}{b_2^2}\lsb
2g_2^2+b_3g_2-g_3b_2-\frac{4}{w}+\frac{1}{w^2}\lb
2g_2^2-b_3g_2+g_3b_2-2b_3g_2\ln w\rb\rsb\\
&&S_3(\xi)=\frac{1}{b_2^3}\lsb -2\lb b_3g_2^2-\fonethird
g_4b_2^2-\fonethird b_4b_2g_2+\frac{2}{3}g_2^3+\fonethird
b_3^2g_2\rb -\frac{2}{w}\lb
g_2g_3b_2-2g_2^3-b_3g_2^2\rb \right. \nonumber \\
&&~~~~~~~~~+\frac{1}{w^2}\lb 4b_3g_2^3\ln w -2\lb
-b_3^2g_2+b_4b_2g_2-b_3g_2^2+g_2^2b_2^2+g_2g_3b_2+2g_2^3\rb\rb
\nonumber \\ &&~~~~~~~~~+ \frac{1}{w^3}\biggl( 2b_3^2g_2 (\ln
w)^2-2(2b_3g_2^2+b_2b_3g_3)\ln w \nonumber\\
&&~~~~~~~~~\left.-2\lb -g_2^2b_2^2-\fonethird
g_4b_2^2-\frac{2}{3}b_4g_2b_2+\fonethird
b_3g_3b_2-g_2g_3b_2+b_3g_2^2+\frac{2}{3}b_3^2g_2-\frac{2}{3}g_2^3\rb
\biggr) \rsb \eea \bea &&S_4(\xi)=\frac{1}{b_2^4}\lsb \frac{2}{3}\lb
\frac{3}{4}\lb
b_3g_4b_2^2+b_5g_2b_2^2+b_4b_2^2g_3-b_3^2g_3b_2-g_5b_2^3\rb
-2b_4b_2g_2^2-\frac{3}{2}b_3b_4g_2b_2+3b_3g_2^3\right.\right.
\nonumber
\\
&&~~~~~~~~~ \left.-\frac{7}{2}b_3g_2b_2
+\frac{11}{4}b_3^2g_2^2-3g_2^2g_3b_2+2g_2g_4b_2^2+g_2^4\rb
\nonumber \\
&&~~~~~~~~~ +\frac{4}{3w}\lb
b_4b_2g_2^2-2g_2^4+b_3g_2g_3b_2-b_3^2g_2^2+3g_2^2g_3b_2-3b_3g_2^3-g_2g_4b_2^2\rb
\nonumber \\
&&~~~~~~~~~+\frac{1}{w^2}\lb
2(-2b_3g_2^3-b_3^2g_2^2+b_3g_2g_3b_2)\right.
\ln w\nonumber\\
&&~~~~~~~~~  \left.
-b_3^3g_2-g_2b_2^3g_3+2b_3g_2b_2(b_4+g_3)-g_3^2b_2^2+4g_2^3b_2^2+b_3g_2^2b_2^2
-5b_3^2g_2^2-b_5g_2b_2^2+4g_2^4+4b_4b_2g_2^2\rb \nonumber \\
&&~~~~~~~~~ +\frac{1}{w^3} \left( 4\left( b_3{g_2 ^2}{b_2^2}+
b_3g_2g_3b_2- {b_3^3}g_{ {2}}+2 b_3{g_2^3}+b_3b_4g_2b_2 y\right) \ln
w-4 {{b_3^2}{g_2^2}(\ln w )^2}\right.
\nonumber \\
&&~~~~~~~~~ +\frac{2}{3} \left(2 b_3g_2g_3b_2-3 b_4{b_2^2}g_{{ 3}}-3
g_2{b_2^3}g_3+6 b_3{g_2^3}-12 {g_{{2} }^3}{b_2^2}-2
g_2g_4{b_2^2}+3 {b_3^2}g_3b_2\right.\nonumber \\
\nonumber \\
&&~~~~~~~~~\left.\left.-4 {g_2^4}-10 b_4b_2{g_2^2}-6
{g_2^2}g_3b_2+10 {b_3^2}{g_2^2}\right)
 \right) \nonumber \\
&&~~~~~~~~~+\frac{1}{w^4} \biggl(-2 {{b_3^3}g_2 (\ln w)^3}+{\left( 6
{b_3^2}{g_2^2}+3 {b_3^2}g_3b_2+2 {b_3^3}g_2\right)(\ln w)^2}
\nonumber
\\ &&~~~~~~~~~ +\frac{2}{3} \left(6
{g_2^3}{b_2^2}+\frac{3}{4} g_{{5} }{b_2^3}+\frac{3}{4}
{g_3^2}{b_2^2}-\frac{9}{4} {b_3^2}g_{{3 }}b_2+\frac{9}{4}
b_4{b_2^2}g_3+4 b_4b_2{g_2}^ {2}+3 {g_2^2}g_3b_2+2
g_2g_4{b_2^2}+\frac{9}{2}
 g_2{b_2^3}g_3 \nonumber \right. \\
&&~~~~~~~~~ \left. +\frac{3}{4} b_5g_2{b_2^2}-\frac{3}{2} b_
3b_4g_2b_2-\frac{7}{2} b_3g_2g_3b_2-\frac{3}{2} b_{{
3}}{g_2^2}{b_2^2}-\frac{3}{4} b_3g_4{b_2^2}+\frac{3}{4} {b
_3^3}g_2-3 b_3{g_2^3}-{\frac{13}{4}} {b_3}^
{2}{g_2^2}+{g_2^4}\right)\nonumber \\
&&~~~~~~~~~ + { \left(-4 b_3{g_2^3}+4 {b_3^3}g_2-4 b_3b_4g_2b_2-6
b_3{g_2^2}{b_2^2}-2 b_3g_4{b_2^2}+2 {b_3^2}{g_2^2}-6 b_3g_2g_{{3}
}b_2\right)\ln w}\biggr)\biggr].\label{s4sol}\eea In eqs.
(\ref{s0sol}-\ref{s4sol}) we have not only the complete five-loop
contribution to $Z$ without having to compute any Feynman diagrams,
but also the LL, $\cdots$, N$^4$LL contribution to $Z$ coming from
all orders of perturbation theory. The LL result of eq.
\refer{s0sol} is consistent with the perturbative calculations of
$Z$ \cite{12,13,14}. We note that it is not a solution to the RG
equation \refer{zrg} when $\beta$ and $\gamma$ are truncated to
lowest order; the solution to this equation involves contributions
from parts of $S_m$ $(m\geq 1)$. The solution \refer{smsol} is
completely determined by the RG functions.

The RG functions \mbeta and \mgamma used to this point are these
associated with the renormalization conditions of eq. \refer{vcd}
and \refer{zcd}. As was pointed out in Ref. \cite{15} and further
developed in Refs. \cite{7,8}, the mass scale $\mu^2$ used in this
on shell scheme is related to the mass scale $\tildemu^2$ used in
the minimal subtraction (MS) scheme through \be
\tildemu^2=\lambda\mu^2\label{mumurel}\ee so that if $\tildebeta$
and $\tildegamma$ are the RG functions in the MS scheme, then \be
\beta(\lambda)=\frac{\tildebeta(\lambda)}{1-\tildebeta(\lambda)/(2\lambda)}\ee
and \be\gamma(\lambda)=
\frac{\tildegamma(\lambda)}{1-\tildebeta(\lambda)/(2\lambda)}.\ee
The functions $\tildebeta$ and $\tildegamma$ are given up to five
loop order in Ref. \cite{16}; this in turn gives \mbeta and \mgamma
to five loop order \cite{7}. Explicit expressions for the MS RG
functions $\tildebeta$ and $\tildegamma$ up to five loop order
appear in Ref. \cite{7}, so that $S_0(\xi),~\cdots,~S_4(\xi)$ can be
given in terms of MS RG functions.  This permits determining $Z$ up
to order N$^4$LL exactly, when using the renormalization condition
of eq. \refer{zcd}. Finding \mcz in some other scheme such as MS
would require explicit computation of the boundary condition
$S_n(0)$ through evaluation of Feynman diagrams in that
renormalization scheme. Recall also that the quantity $L$ in the MS
scheme has explicit dependence on $\lambda$.

We now consider massless scalar electrodynamics.

\section{$Z(\phi)$ in massless scalar
electrodynamics}\label{sechm}

In this section we examine massless scalar electrodynamics, whose
classical action is  \be S_{\mbox{cl}}=\int\dfx\lcb\fhalf\lsb (
\dsubmu+ie\casubmu)\phi^*\rsb \lsb (
\dsupmu-ie\casupmu)\phi\rsb-\frac{1}{4}(\dsubmu A_{\nu}-\dsubnu
\casubmu)^2-\frac{\lambda}{4!}(\phi^*\phi)^2 \rcb.\ee This is
dependent on two coupling constants, \mlambda and $\alpha\equiv
e^2$.

The effective action in the presence of a background field $\phi$
and a background gauge field $A_{\mu}$ can be expanded in powers of
derivatives of these background fields leading to \be
\calcl_{\mbox{eff}}=
-V(\phi)-\frac{1}{4}H(\phi)(F_{\mu\nu})^2+\fhalf
Z(\phi)|\dsubmu\phi-ie\casubmu \phi|^2+\cdots, \ee  where all fields
are evaluated at some fixed point. The form of the dependence of
$\calcl_{\mbox{eff}}$ on \mphi and $A_{\mu}$ is restricted by gauge
invariance.

We again focus on the function $Z(\phi)$ associated with kinetic
term for the scalar field. Once more, we apply an ``on shell''
renormalization condition \cite{1} \be Z(\phi^*\phi=\mu^2)=1.
\label{zmsqedcd}\ee The function \mcz can be evaluated
perturbatively, leading to a general expansion \be
Z(\lambda,\alpha,\phi,\mu)=\sum_{n=0}^{\infty}\sum_{k=0}^{\infty}\sum_{r=0}^{n+k}
T_{n+k-r,r,k}\lambda^{n+k-r}\alpha^{r}L^k,\label{zgf}\ee where
$L=\ln(\phi^*\phi/\mu^2)$ with $\mu^2$ again being the
renormalization scalar parameter. The RG equation is now
 \be \lsb \mu\frac{\partial}{\partial
\mu}+\beta^{\lambda}(\lambda,\alpha) \frac{\partial}{\partial
\lambda}+ \beta^{\alpha}(\lambda,\alpha) \frac{\partial}{\partial
\alpha}+\gamma(\lambda,\alpha)\phi\frac{\partial}{\partial\phi}+
2\gamma(\lambda,\alpha)\rsb Z =0\label{zrgmsqed}\ee where  \bea
\beta^{\lambda}(\lambda,\alpha)&\equiv&\mu\frac{\partial\lambda}{\partial
\mu}=\sum_{n=2}^{\infty}\beta_n^{\lambda},\qquad \beta_n^{\lambda}=
\sum_{r=0}^nb_{n-r,r}^{\lambda}\lambda^{n-r}\alpha^r\,,\\
\beta^{\alpha}(\lambda,\alpha)&\equiv&\mu\frac{\partial\alpha}{\partial
\mu}=\sum_{n=2}^{\infty}\beta_n^{\alpha},\qquad \beta_n^{\alpha}=
\sum_{r=0}^nb_{n-r,r}^{\alpha}\lambda^{n-r}\alpha^r\,,\\
\mbox{and
}\gamma(\lambda,\alpha)&\equiv&\frac{\mu}{\phi}\frac{\partial\phi}{\partial
\mu}=\sum_{n=1}^{\infty}\gamma_n,\qquad \gamma_n=
\sum_{r=0}^ng_{n-r,r}\lambda^{n-r}\alpha^r.\eea (The product
$eA_{\mu}$ is not renormalized \cite{17} with suitable gauge fixing.
)

The kinetic term \mcz of eq. \refer{zgf} depends now on two
couplings $\lambda$ and $\alpha$; this is what makes the discussion
of \mcz in massless scalar electrodynamics more complicated than in
the pure massless $\phi_4^4$ model, where $Z$ in eq. \refer{zexp}
depends on only a single coupling $\lambda$. It turns out to be
advantageous to define \cite{8} \be
P_n^k(\lambda,\alpha)=\sum_{r=0}^n T_{n-r,r,k}\lambda^{n-r}\alpha^r
\label{pdef}\ee so that by eq. \refer{zgf} \be
Z(\phi)=\sum_{n=0}^{\infty}\sum_{k=0}^{n}P_n^k(\lambda,\alpha) L^k
\label{zreorgms}. \ee Just as $S_m(\lambda L)$ of eq. \refer{sdef}
is identified with the N$^m$LL contribution to \mcz in the massless
$\phi_4^4$ model, so also we find that the N$^n$LL contribution to
\mcz in massless scalar electrodynamics is given by \be
Z_{N^nLL}=\sum_{k=0}^{\infty}P_{k+n}^kL^k.\label{zprel}\ee With the
renormalization condition of eq. \refer{zmsqedcd}, we have \be
P_n^0=\delta_{n0}. \label{pbc}\ee

Substitution of eq. \refer{zreorgms} into \refer{zrgmsqed} results
in \be \sum_{n=0}^{\infty}\sum_{k=0}^n\lsb 2k(-1+\sum_{m=1}^{\infty}
\gamma_m)P_n^kL^{k-1}+\sum_{m=2}^{\infty}\lb
\beta_m^{\lambda}\frac{\partial}{\partial\lambda}+\beta_m^{\alpha}
\frac{\partial}{\partial \alpha}\rb
P_{n}^kL^k+2\sum_{m=1}^{\infty}\gamma_mP_n^kL^k\rsb=0.\label{zsubms}\ee
Each of the quantities $P_n^k$, $\beta_n^{\lambda}$,
$\beta_n^{\alpha}$ and $\gamma_n$ are of the form $\displaystyle
\sum_{r=0}^nC_{n-r,r}\lambda^{n-r}\alpha^r$, i.e., they are
polynomials of degree \mn in \mlambda and $\alpha$. By having each
term of order $k$ in \mcl and of order \mn in \mlambda and \malpha
in eq. \refer{zsubms} equaling zero, we obtain a series of coupled
partial differential equations with the boundary condition of eq.
\refer{pbc} that can be solved for each of the $P_n^k$ in turn. For
example, if in equation \refer{zsubms} we were to consider the term
of zeroth order in \mcl and first order in the coupling, we have \be
-2P_1^1+2\gamma_1P_0^0=0. \ee This, combined with eq. \refer{pbc},
determines $P_1^1$. In general, if in eq. \refer{zsubms} we examine
terms of order \mn in \mcl and order $(n+1)$ in the coupling, then
\be P_{n+1}^{n+1}=\frac{1}{n+1}\lsb\fhalf
\lb\beta_2^{\lambda}\frac{\partial}{\partial
\lambda}+\beta^{\alpha}_2\frac{\partial}{\partial
\alpha}\rb+\gamma_1\rsb P_{n}^{n} \label{pnp1np1}.\ee This fixes all
the LL contributions to $Z_{LL}$ of eq. \refer{zprel} in terms of
$\beta_2^{\lambda}$, $\beta_2^{\alpha}$ and $\gamma_1$. One could
obtain each of the coefficients $T_{n-r,r,n}$ contributing to
$P_n^n$ in eq. \refer{pdef} in terms of $b_{2-r,r}^{\lambda}$,
$b_{2-r,r}^{\alpha}$ and $g_{1-r,r}$ directly from eq.
\refer{pnp1np1}.

The coefficients $P^{n}_{n+1}$ that contribute to $Z_{NLL}$ can also
be found from the RG equation of eq. \refer{zsubms}. To do this we
consider these terms in eq. \refer{zsubms} that are of order \mn in
\mcl and order $(n+2)$ in the couplings, \be \lb
P^{n+1}_{n+2}-\gamma_1P^{n+1}_{n+1}\rb=\frac{1}{n+1}\lcb
\lsb\fhalf\lb\beta_2^{\lambda}\frac{\partial}{\partial
\lambda}+\beta^{\alpha}_2\frac{\partial}{\partial
\alpha}\rb+\gamma_1\rsb P^{n}_{n+1}
+\lsb\fhalf\lb\beta_3^{\lambda}\frac{\partial}{\partial
\lambda}+\beta^{\alpha}_3\frac{\partial}{\partial
\alpha}\rb+\gamma_2\rsb P_{n}^{n}\rcb \label{pnp1np2}.\ee

Having determined each of the coefficients $T_{n-r,r,n}$ that
contribute to $P^n_n$ from eq. \refer{pnp1np1}, we can now use eq.
\refer{pnp1np2} to fix the coefficients $T_{n+1-r,r,n}$ that go into
$P^n_{n+1}$ provided $\beta_3^{\lambda}$, $\beta_3^{\alpha}$ and
$\gamma_2$ are known. The boundary condition $P^0_1=0$ which follows
from eq. \refer{pbc} is also employed.

As in Ref. \cite{8}, a variant of the method of characteristics can
be used to find a closed form expression for the sums occurring in
$Z_{LL}$ and $Z_{NLL}$. We start by defining \be
W_{n+k}^n(\barlambda(t), \baralpha(t),t)=\exp\lsb
2\int_0^t\gamma_1(\barlambda(\tau),\baralpha(\tau))d\tau\rsb
P_{n+k}^n(\barlambda(t),\baralpha(t))\label{wdef}\ee where the
characteristic functions $\barlambda(t)$ and $\baralpha(t)$ satisfy
\bea &&
\frac{d\barlambda(t)}{dt}=\beta_2^{\lambda}(\barlambda(t),\baralpha(t))\,,
\qquad \barlambda(0)=\lambda,\label{chalam}\\
&&
\frac{d\baralpha(t)}{dt}=\beta_2^{\alpha}(\barlambda(t),\baralpha(t))\,,
\qquad \baralpha(0)=\alpha.\label{chal}\eea The forms of
$\beta_2^{\lambda}$ and $\beta_2^{\alpha}$ appear in Ref. \cite{1},
as well as the solution for $\barlambda(t)$ and $\baralpha(t)$.
Differentiating \refer{wdef} with respect to $t$ gives \be
\frac{d}{dt}W_{n+k}^n(\barlambda(t),\baralpha(t),t)=\lb
\beta_2^{\lambda}(\barlambda,\baralpha)\frac{\partial}{\partial
\barlambda}+
\beta_2^{\alpha}(\barlambda,\baralpha)\frac{\partial}{\partial
\baralpha}+2\gamma_1(\barlambda,\baralpha)\rb
W_{n+k}^n(\barlambda(t),\baralpha(t),t).\label{wdif} \ee Together,
equations \refer{pnp1np1}, \refer{wdef} and \refer{wdif} show that
\be
W_{n+1}^{n+1}=\frac{1}{2(n+1)}\frac{d}{dt}W^n_n=\frac{1}{2^{n+1}(n+1)!}
\frac{d^{n+1}}{dt^{n+1}}W_0^0. \label{wrel}\ee If now \be
\barcz_{LL}(t)=\sum_{n=0}^{\infty}W_n^n(\barlambda(t),\baralpha(t),t)
\barcl^n\label{barczdef}\ee where \be
\barcl=\ln\lb\frac{\phi^*\phi}{\barmu^2(t)}\rb\ee with \be
\frac{d\barmu(t)}{dt}=\barmu(t)\,,\qquad \barmu(0)=\mu,\qquad\mbox{
so that }\barmu(t)=\mu e^t,\label{chmu}\ee then by \refer{wrel} \be
\barcz_{LL}(t)=\sum_{n=0}^{\infty}\frac{1}{n!}\lb\frac{\barcl}{2}
\rb^n\lb\frac{d}{dt}\rb^nW_0^0(\barlambda(t),\baralpha(t),t)=
W_0^0\lb\barlambda(t+\frac{\barcl}{2}),
\baralpha(t+\frac{\barcl}{2}),t+\frac{\barcl}{2}\rb.\label{barczsol}\ee
Furthermore, eqs. (\ref{wdef})-(\ref{chal}),
(\ref{barczdef})-(\ref{chmu}) show that \be \barcz_{LL}(0)=Z_{LL}\ee
and so by eqs. \refer{pbc} and \refer{barczsol}, \be Z_{LL}=
\exp\lsb 2\int_0^{L/2}
\gamma_1(\barlambda(\tau),\baralpha(\tau))d\tau\rsb.
\label{llsol}\ee Since $L=0$ when $\phi^*\phi=\mu^2$, the
renormalization condition of eq. \refer{zmsqedcd} is satisfied.

Next we take \be
\barcz_{N^mLL}(t)=\sum_{n=0}^{\infty}W_{n+m}^n(\barlambda(t),
\baralpha(t),t)\barcl^n \label{barcznlldef}\ee so that \be
\barcz_{N^mLL}(0)=Z_{N^mLL}. \label{barczczrel} \ee By eqs.
\refer{pnp1np2} and \refer{wdef}, we find that \be
W^n_{n+1}=\frac{1}{2n}\lsb
 \lb \beta_2^{\lambda}\frac{\partial}{\partial\barlambda}+ \beta_2^{\alpha}
 \frac{\partial}{\partial\baralpha}+2\gamma_1\rb W_n^{n-1}+\lb
\beta^{\lambda}_3\frac{\partial}{\partial\barlambda}+\beta^{\alpha}_3
\frac{\partial}{\partial\baralpha} +2\gamma_2\rb W_{n-1}^{n-1}\rsb
+\gamma_1W^n_n.\ee  Using eqs. \refer{pnp1np1} and\refer{wdif}, this
equation becomes \bea W^n_{n+1}\aeq\frac{1}{2n}\lsb
 \lb \beta_2^{\lambda}\frac{\partial}{\partial\barlambda}+ \beta_2^{\alpha}
 \frac{\partial}{\partial\baralpha}+2\gamma_1\rb W_n^{n-1}\right.\nonumber\\
&&~~~~~\left.+\lb\lb
\beta^{\lambda}_3\frac{\partial}{\partial\barlambda}+\beta^{\alpha}_3
\frac{\partial}{\partial\baralpha} +2\gamma_2\rb +\gamma_1 \lb
\beta^{\lambda}_2\frac{\partial}{\partial\barlambda}+\beta^{\alpha}_2
\frac{\partial}{\partial\baralpha} +2\gamma_1\rb\rb W_{n-1}^{n-1}\rsb \label{odefp}\\
&\equiv& \frac{1}{2n}\lsb
\frac{d}{dt}W^{n-1}_n+\calcd(t)W^{n-1}_{n-1}\rsb. \label{odef}\eea
Iterating eq. \refer{odef} and again using eq. \refer{wrel}, we find
that \be W_{n+1}^n =\frac{1}{2^nn!}\lsb
\frac{d^n}{dt^n}W_1^0+\calcd(t)\frac{d^{n-1}}{dt^{n-1}}W_0^0+\cdots+
\frac{d^{n-2}}{dt^{n-2}}\lb\calcd(t)\frac{d}{dt}W_0^0\rb +
\frac{d^{n-1}}{dt^{n-1}}\lb\calcd(t)W_0^0\rb\rsb.\label{wid} \ee
Using identity \bea && ~ \frac{d^n}{dt^n}\lb
fg\rb+\frac{d^{n-1}}{dt^{n-1}}\lb f\frac{dg}{dt}\rb +\cdots
+\frac{d}{dt}\lb
f\frac{d^{n-1}g}{dt^{n-1}}\rb+f\frac{d^ng}{dt^n}\nonumber\\
\aeq \frac{d^{n+1}}{dt^{n+1}}\lb \phi g\rb
-\phi\frac{d^{n+1}}{dt^{n+1}}g\qquad (\phi'=f),\label{iden}\eea eq.
\refer{wid} reduces to \be W_{n+1}^n = \frac{1}{2^nn!}\lsb
\frac{d^n}{dt^n}\lb W_1^0 +\tilde{\calcd}(t)W_0^0\rb
-\tilde{\calcd}(t)\frac{d^n}{dt^n}W_0^0\rsb ,\label{todef}\ee where
$d\tilde{\calcd}(t)/dt\equiv\calcd(t)$, so that \be
\tilde{\calcd}(t)W^0_0(\barlambda(t),\baralpha(t),t)=\lb\int_{t_0}^t\lsb
\beta_3^{\lambda}\lb\barlambda(\tau),\baralpha(\tau)\rb
\frac{\partial}{\partial\barlambda(t)}+\cdots
+2\gamma_1^2\lb\barlambda(\tau),\baralpha(\tau)\rb\rsb d\tau\rb
W^0_0\lb\barlambda(t),\baralpha(t),t\rb\label{tddef}\ee and \bea
&&\frac{d}{dt}\lb\tilde{\calcd}(t)W_0^0(\barlambda(t),\baralpha(t),t)\rb\nonumber\\
&=&\lsb
\beta_3^{\lambda}(\barlambda(t),\baralpha(t))\frac{\partial}{\partial\barlambda(t)}
+\cdots +2\gamma_1^2(\barlambda(t),\baralpha(t))\rsb
W_0^0(\barlambda(t),\baralpha(t),t)\nonumber\\
&&+\int_0^t\lsb
\beta_3^{\lambda}(\barlambda(\tau),\baralpha(\tau))\frac{\partial}{\partial
\barlambda(t)}+\cdots+2\gamma_1^2(\barlambda(\tau),\baralpha(\tau))\rsb
d\tau\frac{d}{dt}W_0^0(\barlambda(t),\baralpha(t),t).\label{dtddef}\eea
Eqs. \refer{tddef} and \refer{dtddef} ensure consistency between
eqs. \refer{wid} and \refer{todef}. It is important to note that in
eqs. \refer{tddef} and \refer{dtddef} $\barlambda$ and $\baralpha$
are evaluated at $t$ when they appear in the arguments of $W_0^0$
and also that derivatives with respect to $\barlambda$ and
$\baralpha$ are also computed with $\barlambda$ and $\baralpha$
evaluated at $t$. In eq. \refer{dtddef}, the ordinary derivative
$d/dt$ acts on $W_0^0(\barlambda(t),\baralpha(t),t)$ prior to
functional derivatives $\partial/\partial\barlambda(t)$ and
$\partial/\partial\baralpha(t)$. The last step to be performed on
the right hand side of eq. \refer{dtddef} is the integration over
$\tau$. Eq. \refer{barcznlldef} now can be written as \be
\barcz_{NLL}(t)=\sum_{n=0}^{\infty}\frac{\barcl^n}{2^nn!}\lsb
\frac{d^n}{dt^n}\lb
W_1^0(\barlambda(t),\baralpha(t),t)+\tilde{\calcd}(t)W^0_0(\barlambda(t),\baralpha(t),t)\rb
-\tilde{\calcd}(t)\frac{d^n}{dt^n}W_0^0(\barlambda(t),\baralpha(t),t)\rsb.
\label{barczres}\ee The sum over \mn in eq. \refer{barczsol} can now
be performed and upon setting $t=0$ and using eq. \refer{barczczrel}
we obtain \be Z_{NLL}(0)=W_1^0\lb
\barlambda\lb\frac{L}{2}\rb,\baralpha\lb\frac{L}{2}\rb,
\frac{L}{2}\rb +\tilde{\calcd}\lb\frac{L}{2}\rb W_0^0\lb
\barlambda\lb\frac{L}{2}\rb,\baralpha\lb\frac{L}{2}\rb,
\frac{L}{2}\rb \ee provided $\tilde{\calcd}(0)=0$ (i.e., we select
$t_0=0$). We thus have a closed form expression for $Z_{NLL}$. The
renormalization condition of eq. \refer{pbc} further reduces it to
\bea Z_{NLL}&=& \tilde{\calcd}\lb\frac{L}{2}\rb \exp\lsb
2\int_0^{L/2}\gamma_1\lb\barlambda(\tau),\baralpha(\tau)\rb
d\tau\rsb\nonumber\\
&=&2\int_0^{L/2}d\tau\lsb\gamma_2(\tau)+
\gamma_1^2(\tau)\rsb\exp\lsb2\int_0^{L/2}\gamma_1(\barlambda(\tau),
\baralpha(\tau))d\tau\rsb.\label{nllsol}\eea

The closed form of $Z_{N^2LL}$ term could be obtained in a similar
manner. In eq. \refer{zsubms}, consider the terms that are of order
$n$ of $L$ and of order $(n+3)$ in the couplings, \bea
P^{n+1}_{n+3}-\lb\gamma_1 P^{n+1}_{n+2}+\gamma_2 P^{n+1}_{n+1}\rb
\aeq\frac{1}{2(n+1)}\lsb \lb
\beta_2^{\lambda}\frac{\partial}{\partial \lambda}
+\beta_2^{\alpha}\frac{\partial}{\partial \alpha}+2\gamma_1\rb
P^n_{n+2}\right.\nonumber\\
&&\left.+\lb \beta_3^{\lambda}\frac{\partial}{\partial \lambda}+
\beta_3^{\alpha}\frac{\partial}{\partial \alpha}+2\gamma_2 \rb
P^n_{n+1} + \lb \beta_4^{\lambda}\frac{\partial}{\partial \lambda} +
\beta_4^{\alpha}\frac{\partial}{\partial \alpha}+2\gamma_3\rb
P^n_n\rsb. \eea Using eqs. \refer{wrel} and \refer{odef}, we obtain
\bea W^{n+1}_{n+3}\aeq\frac{1}{2(n+1)}\lsb \frac{d}{dt}W^n_{n+2}
+\lb \gamma_1\lb\beta^{\lambda}_2\frac{\partial}{\partial
\barlambda}+\beta^{\alpha}_2\frac{\partial}{\partial
\baralpha}+2\gamma_1\rb+\beta_3^{\lambda}\frac{\partial}{\partial
\barlambda}+\beta_3^{\alpha}\frac{\partial}{\partial
\baralpha}+2\gamma_2\rb W^n_{n+1}\right.\nonumber\\
&&\qquad\qquad+\lb
(\gamma_2+\gamma_1^2)\lb\beta^{\lambda}_2\frac{\partial}{\partial
\barlambda}+\beta^{\alpha}_2\frac{\partial}{\partial
\baralpha}+2\gamma_1\rb+\gamma_1\lb\beta^{\lambda}_3\frac{\partial}{\partial
\barlambda}+\beta^{\alpha}_3\frac{\partial}{\partial
\baralpha}+2\gamma_2\rb\right.
\nonumber\\
&&\qquad\qquad+\left.\left.\beta_4^{\lambda}\frac{\partial}{\partial
\barlambda} +\beta_4^{\alpha}\frac{\partial}{\partial
\baralpha}+2\gamma_3\rb W_n^n\rsb\\
&\equiv& \frac{1}{2(n+1)}\lsb
\frac{d}{dt}W^n_{n+2}+\calcd_1(t)W^n_{n+1}+\calcd_0(t)W^n_n\rsb.
\eea Iterating this equation $n$ times and again using \refer{wrel}
and \refer{todef}, it is found \bea
W^{n+1}_{n+3}\aeq\frac{1}{2^{n+1}(n+1)!} \lsb
\frac{d^{n+1}}{dt^{n+1}}W^0_2+\calcd_1(t)\frac{d^n}{dt^n}\lb
W^0_1+\tilde{\calcd}(t)W^0_0\rb
-\calcd_1(t)\tilde{\calcd}(t)\frac{d^n}{dt^n}W_0^0\right.\nonumber\\
&&\qquad\qquad\qquad+
\frac{d}{dt}\calcd_1(t)\frac{d^{n-1}}{dt^{n-1}}\lb
W^0_1+\tilde{\calcd}(t)W^0_0\rb
-\frac{d}{dt}\lb\calcd_1(t)\tilde{\calcd}(t)\rb\frac{d^{n-1}}{dt^{n-1}}
W_0^0\nonumber\\
&&\qquad\qquad\qquad+\cdots\nonumber\\
&&\qquad\qquad\qquad+\frac{d^n}{dt^n}\lsb\calcd_1(t)\lb
W^0_1+\tilde{\calcd}(t)W^0_0\rb\rsb
-\frac{d^n}{dt^n}\lb\calcd_1(t)\tilde{\calcd}(t)W_0^0\rb\nonumber\\
&&\qquad\qquad\qquad\left.+\calcd_0(t)\frac{d^n}{dt^n}W_0^0+
\frac{d}{dt}\calcd_0(t)\frac{d^{n-1}}{dt^{n-1}}W_0^0 +\cdots
+\frac{d^n}{dt^n}\calcd_0(t)W_0^0\rsb.\eea Using eqs. \refer{iden}
and \refer{barcznlldef} and the boundary condition \refer{pbc}, it
is found that \bea
Z_{N^2LL}=\barcz_{N^2LL}(0)\aeq\lb\tilde{\calcd_1}\lb\frac{L}{2}\rb
\tilde{\calcd}\lb\frac{L}{2}\rb -\tilde{\calcd}_{10}
\lb\frac{L}{2}\rb\rb W_0^0 \lb
\barlambda\lb\frac{L}{2}\rb,\baralpha\lb\frac{L}{2}\rb,
\frac{L}{2}\rb \nonumber\\ \aeq \lb\tilde{\calcd_1}\lb\frac{L}{2}\rb
\tilde{\calcd}\lb\frac{L}{2}\rb -\tilde{\calcd}_{10}
\lb\frac{L}{2}\rb+\tilde{\calcd}_{0} \lb\frac{L}{2}\rb\rb \exp\lsb
2\int_0^{L/2}\gamma_1\lb\barlambda(\tau),\baralpha(\tau)\rb
d\tau\rsb \label{znnllres}\eea where
$d\tilde{\calcd}_0(t)/dt=\calcd_0(t)$,
$d\tilde{\calcd_1}(t)/dt=\calcd_1(t)$,
$d\tilde{\calcd}_{10}(t)/dt=\calcd_1(t)\tilde{\calcd}(t)$ with
equations analogous to eqs. \refer{tddef} and \refer{dtddef} being
satisfied and we have set $\tilde{\calcd}_0(0)=0$,
$\tilde{\calcd}_1(0)=0$ and $\tilde{\calcd}_{10}(0)=0$.

For $Z_{N^mLL},(m>2)$, we easily see from the derivation of
$Z_{LL}$, $Z_{NLL}$ and $Z_{N^2LL}$ in eqs. \refer{llsol},
\refer{nllsol} and \refer{znnllres}, that every $Z_{N^mLL}$ is in
the form of some operator $\tilde{\calcd}_m$ applied to $W_0^0$,
i.e., \be Z_{N^mLL}=\tilde{\calcd}_m\lb\frac{L}{2}\rb \exp\lsb
2\int_0^{L/2}\gamma_1\lb\barlambda(\tau),\baralpha(\tau)\rb
d\tau\rsb.\label{zmres}\ee The \mbeta and \mgamma functions to order
$\beta_{m+2}^{\lambda}$, $\beta_{m+2}^{\alpha}$ and $\gamma_{m+1}$
uniquely determine the operator $\tilde{\calcd}_m$, which in turn
fixes $Z(\phi)$ to order N$^m$LL.

From eq. \refer{llsol} it follows that $Z_{LL}=1$ if
$\phi^2=\mu^2~(L=0)$ while from eqs.
(\ref{nllsol},\ref{znnllres},\ref{zmres}) we see that
$Z_{N^pLL}=0~(p>0)$ in this limit. This is consistent with the
renormalization condition of eq. \refer{zcd}. If we take $\mu=v$,
then it is apparent that $Z(v)$ in eq. \refer{massdef} is just equal
to one. The RG function used in deriving eqs. \refer{llsol},
\refer{nllsol}, \refer{znnllres} and \refer{zmres} are those
associated with the ``on shell'' renormalization conditions of eq.
\refer{vcd} and \refer{zcd} or \refer{zmsqedcd}.

\section{Discussion}

We have demonstrated that the kinetic term in the derivative
expansion of the effective Lagrangian can be determined entirely by
the ``on shell'' RG functions in the massless $\phi_4^4$ model. The
LL , NLL, N$^2$LL have similarly been computed in massless scalar
electrodynamics. This is a necessary ingredient in determining the
radiatively generated mass in these models, as is shown by eq.
\refer{massdef}.

There are several avenues that should be further explored. For
example, we should show that terms in the derivative expansion of
the effective Lagrangian beyond the kinetic term for the scalar
field are expressible in terms of the RG functions. We should also
consider the effect of inserting a mass for the scalar fields in the
classical Lagrangian considered here.

Possibly the most important question is to see if the techniques
employed here and in Ref. \cite{7,8} can be used to compute more
precisely the Higgs mass in the massless standard model, thereby
improving on the estimates of Refs. \cite{5,6}. This is currently
being undertaken \cite{22}.
\\
\noindent {\bf Acknowledgements}

D.G.C. McKeon would like to thank USP for its hospitality while some
of this work was done. Roger Macleod had a useful comment.

\end{document}